\documentclass[12pt]{iopart}
\usepackage{epsfig} 
\begin{document}

\title[Energy dependence of {\rm K}$^0_{\rm S}$ and 
hyperon production at CERN SPS]{Energy dependence 
of K$^0_{\rm S}$ and hyperon production at CERN SPS}

\author{D~Elia for the NA57 Collaboration\footnote[1]{For
the full author list see Appendix ``Collaborations'' in this
volume.}
}

\address{INFN Sezione di Bari, Bari, Italy} 

%\address{$^{a}$ Physics Department, University of Athens, Athens, Greece} 
%\address{$^{b}$ Dipartimento IA di Fisica dell'Universit{\`a} e del Politecnico and INFN, Bari, Italy} 
%\address{$^{c}$ Fysisk Institutt, Universitetet i Bergen, Bergen, Norway} 
%\address{$^{d}$ H{\o}gskolen i Bergen, Bergen, Norway} 
%\address{$^{e}$ School of Physics and Astronomy, University of Birmingham, Birmingham, UK} 
%\address{$^{f}$ Comenius University, Bratislava, Slovakia} 
%\address{$^{g}$ University of Catania and INFN, Catania, Italy} 
%\address{$^{h}$ CERN, European Laboratory for Particle Physics, Geneva, Switzerland} 
%\address{$^{i}$ Institute of Experimental Physics, Slovak Academy of Science, Ko\v{s}ice, Slovakia} 
%\address{$^{j}$ P.J. \v{S}af\'{a}rik University, Ko\v{s}ice, Slovakia} 
%\address{$^{k}$ Fysisk Institutt, Universitetet i Oslo, Oslo, Norway} 
%\address{$^{l}$ University of Padua and INFN, Padua, Italy} 
%\address{$^{m}$ Coll\`ege de France, Paris, France} 
%\address{$^{n}$ Institute of Physics, Academy of Sciences of the CR, Prague, Czech Republic} 
%\address{$^{o}$ University ``La Sapienza'' and INFN, Rome, Italy} 
%\address{$^{p}$ Dip. di Scienze Fisiche ``E.R. Caianiello'' dell'Universit{\`a} and INFN, Salerno, Italy} 
%\address{$^{q}$ State University of St. Petersburg, St. Petersburg, Russia} 
%\address{$^{r}$ Institut de Recherches Subatomique, IN2P3/ULP, Strasbourg, France} 
%\address{$^{s}$ Utrecht University and NIKHEF, Utrecht, The Netherlands} 

\begin{abstract}
Recent results on K$^0_{\rm S}$ and hyperon production in Pb-Pb 
collisions at 40 and 158 $A$ GeV/$c$ beam momentum
from the NA57 experiment at CERN SPS are presented.
Yields and ratios are compared with those
measured by the NA49 experiment, where available. The centrality
dependence of the yields and a comparison with
the higher collision energy data from RHIC are 
discussed.

\end{abstract}




%\ead{Domenico.Elia@ba.infn.it}
%\maketitle

\section{Introduction}

The main physics aim of the NA57 experiment~\cite{NA57prop} 
at CERN SPS is the study of the onset of the multi-strange baryon 
and antibaryon enhancements in Pb-Pb collisions
with respect to proton-induced
reactions.
The enhancement effect was first observed by WA97
at 158 $A$ GeV/$c$ beam momentum~\cite{And99}:
NA57 has extended the WA97 measurements
over a wider centrality range and
to lower beam momentum.
The experimental apparatus, described in details 
elsewhere~\cite{Man99},
detects strange and multi-strange
hyperons by reconstructing their weak decays
into final states containing charged particles only.
%Tracks are measured in the silicon telescope, a 30 cm length
%array of pixel detector planes
%with 5 $\times$ 5 cm$^2$ cross section.
%Additional pixel planes and
%double-sided silicon microstrip detectors, placed
%behind the telescope, are used as a lever arm
%to improve the momentum resolution for fast tracks.
%The centrality trigger, based on a scintillator
%petal system placed 10 cm downstream of the target,
The centrality trigger 
selects the most central about 60\% of the inelastic
cross section for Pb-Pb collisions.
The centrality of the collision is controlled
through accurate analysis of the charged particle 
multiplicity sampled at central rapidity by two stations 
of silicon strip detectors.
%The apparatus is placed inside the
%1.4 Tesla field of the GOLIATH magnet.

\section{Results and discussion}

The particle selection procedure is
based on geometrical and kinematical cuts and
allows the extraction of clean
signals with negligible background~\cite{Man2002}.
The collision centrality is expressed as the number
of wounded nucleons computed from 
the multiplicity distribution and the measured
trigger cross section via the Glauber model,
according to the procedure described in \cite{Ant2000}.
\\
The double-differential distribution in rapidity $y$ and
transverse mass $m_{\rm T}$
(${d^2N}/{dm_{\rm T} dy}$)
for each particle type has been fitted, with the corresponding
inverse slope
parameters ($T_{app}$) extracted as reported
in \cite{BrunoQM04}.
Yields have then been
calculated as the number of particles per event extrapolated
to a common phase space
window, covering full $p_{\rm T}$ and one unit of rapidity
around midrapidity:
%using the maximum likelihood
%method, according to the following parametrization:
%
%\begin{equation}
%\frac{d^2N}{dm_T dy}=f(y) \hspace{1mm} m_T \exp\left(-\frac{m_T}{T}\right)
%\hspace{65mm}.     
%\label{eqmtfit}
%\end{equation}

%\noindent
%We assume a flat rapidity distribution
%within our acceptance region around central rapidity and leave
%the inverse slope $T$ as a free parameter.

\begin{equation}
Y = \int_{m}^{\infty} dm_{\rm T} \int_{y_{cm}-0.5}^{y_{cm}+0.5} dy \frac{d^2N}{dm_{\rm T} dy}
\hspace{64mm}.     
\label{eqyield}
\end{equation}

In the following of this paper, after a comparison 
with the results from the NA49 experiment,
we show the dependence of the K$^0_{\rm S}$, $\Lambda$,
$\Xi$ and $\Omega$ yields on centrality and energy
in Pb-Pb collisions. 
A comparison with data from
Au-Au collisions at RHIC is included.
The analysis of the
transverse mass spectra and the energy dependence of
the hyperon enhancements are discussed in detail in the contribution
to this conference by G E Bruno~\cite{BrunoQM04}.

\subsection{Comparison with NA49}
%\label{int}

We have compared hyperon yields and ratios at both
40 and 158 $A$ GeV/$c$ beam momentum
with the available published
results from corresponding measurements performed 
by the NA49 Collaboration~\cite{NA49ref}.
%by the NA49 Collaboration~\cite{NA49y1,NA49y2,NA49y3,NA49y4,NA49y5}.
For this comparison we have restricted our data to
centrality ranges corresponding to those of NA49; the 
K$^0_{\rm S}$ yields for NA49 have been extracted from the
published yields for charged kaons.
\par
\begin{figure}[h]
\centering
\includegraphics[scale=0.42]{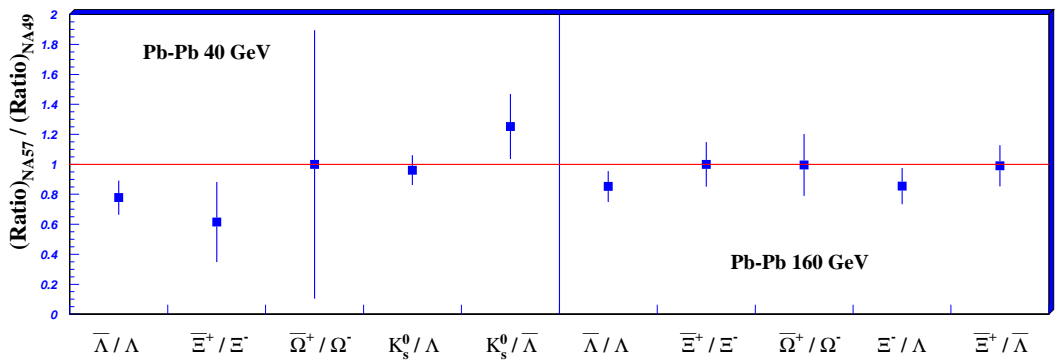}
\caption{\label{figratcompna49} Comparison of particle ratios
measured by NA57 and NA49 experiments.} 
\end{figure}

The NA49 yields are systematically lower than those
measured by NA57 by about 30\%,
both in the 40 and 158 $A$ GeV/$c$ data.
Investigations are ongoing to find out the source
of this discrepancy.
Neverthless, as shown in Figure \ref{figratcompna49}, 
the systematics on
the absolute yields cancel out when calculating particle ratios.

\subsection{Energy and centrality dependence of the yields}

\begin{figure}[htb]
\hspace{15mm} 
\includegraphics[scale=0.47]{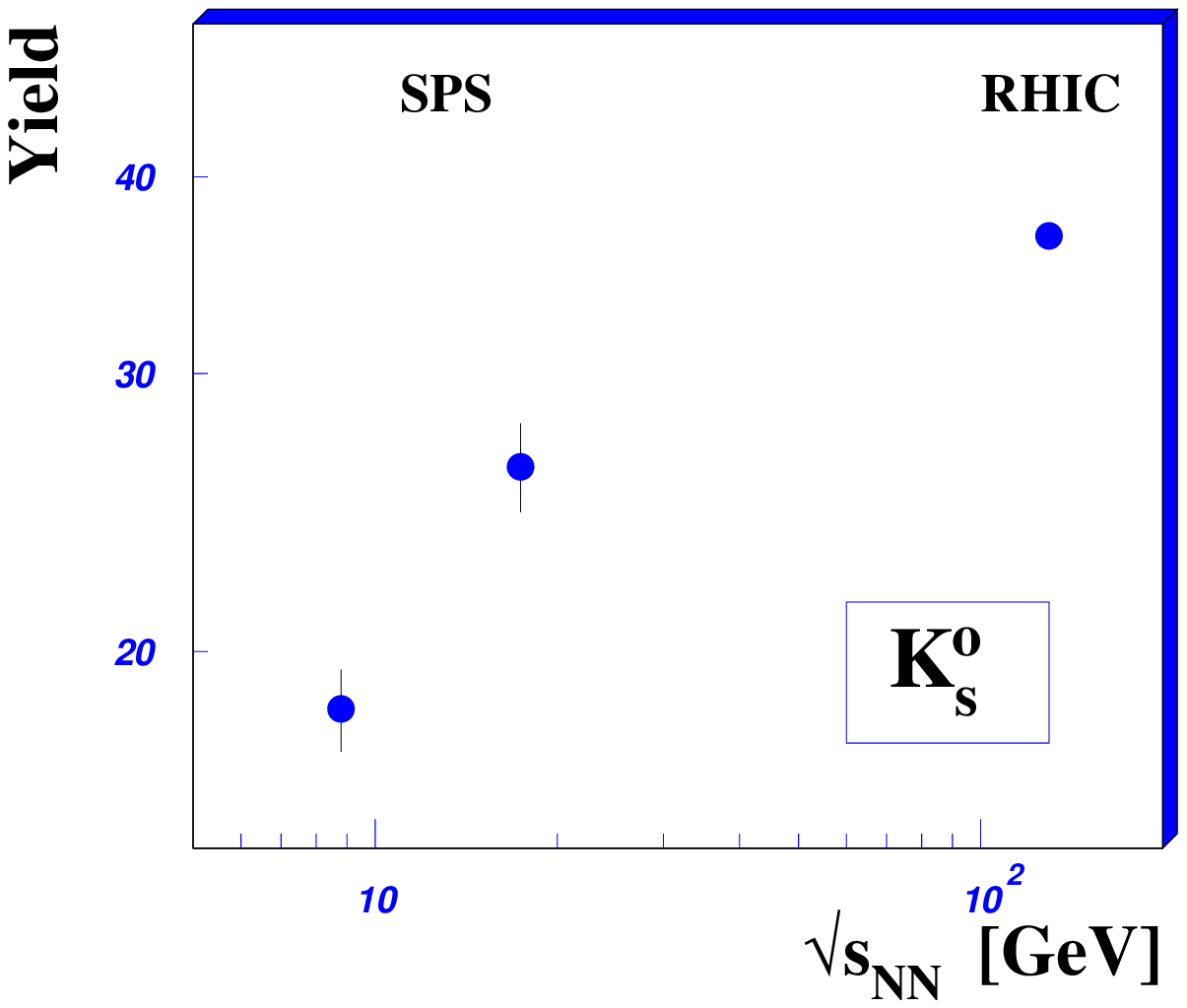}
\hspace{\fill}
\begin{minipage}[t]{110mm}
\includegraphics[scale=0.47]{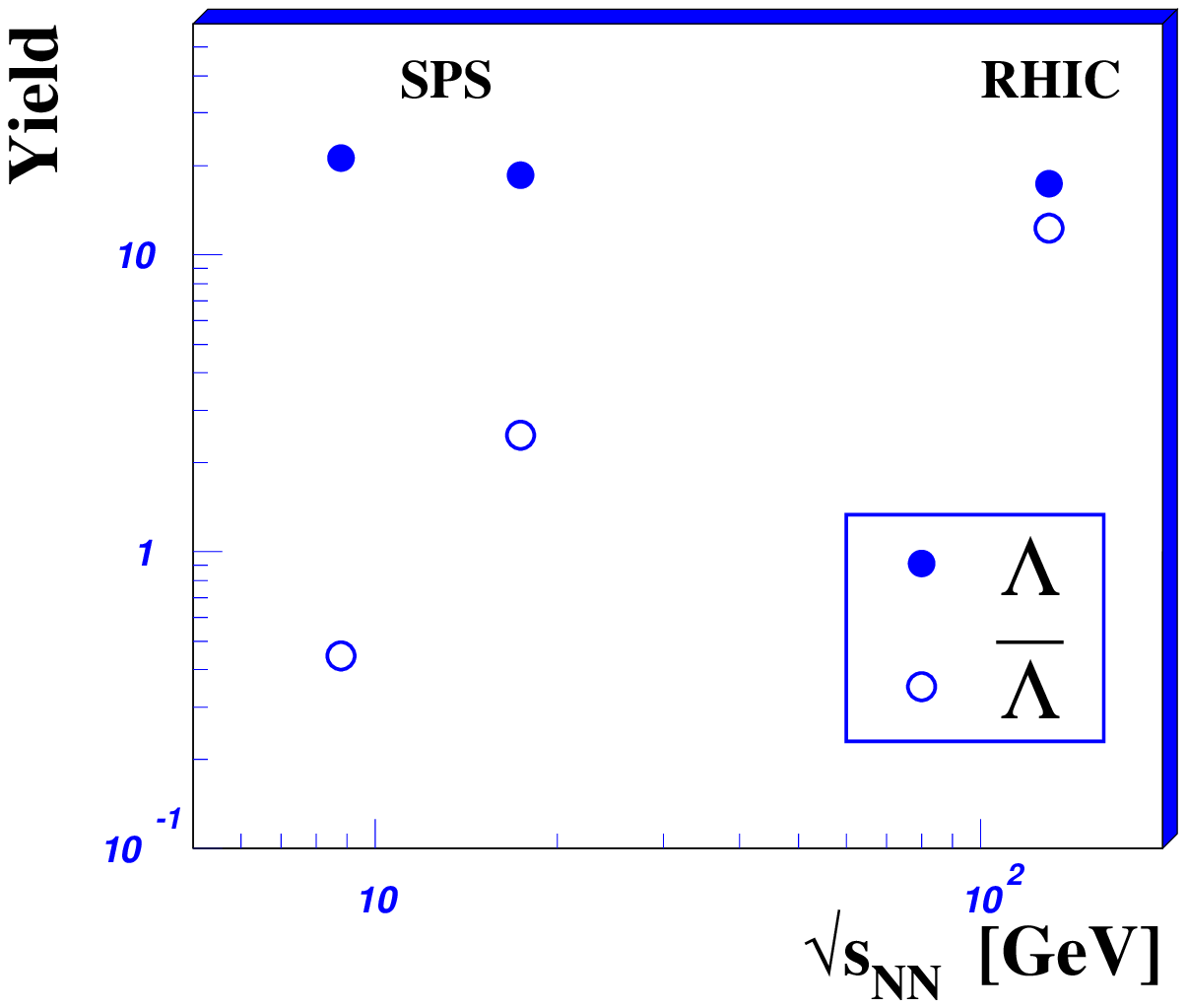}
\end{minipage}
\par
\hspace{15mm} 
\includegraphics[scale=0.47]{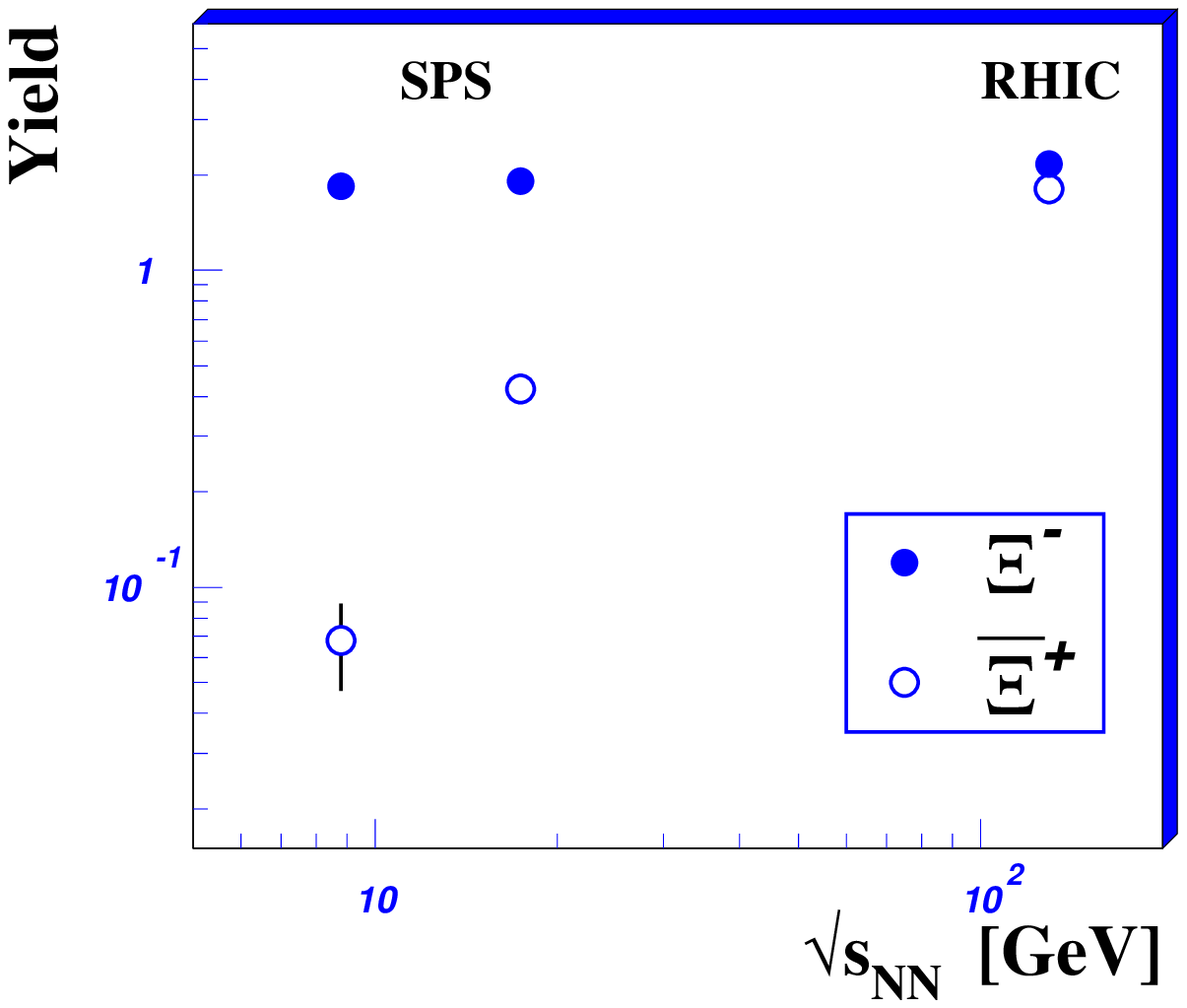}
\hspace{\fill}
\begin{minipage}[t]{110mm}
\includegraphics[scale=0.47]{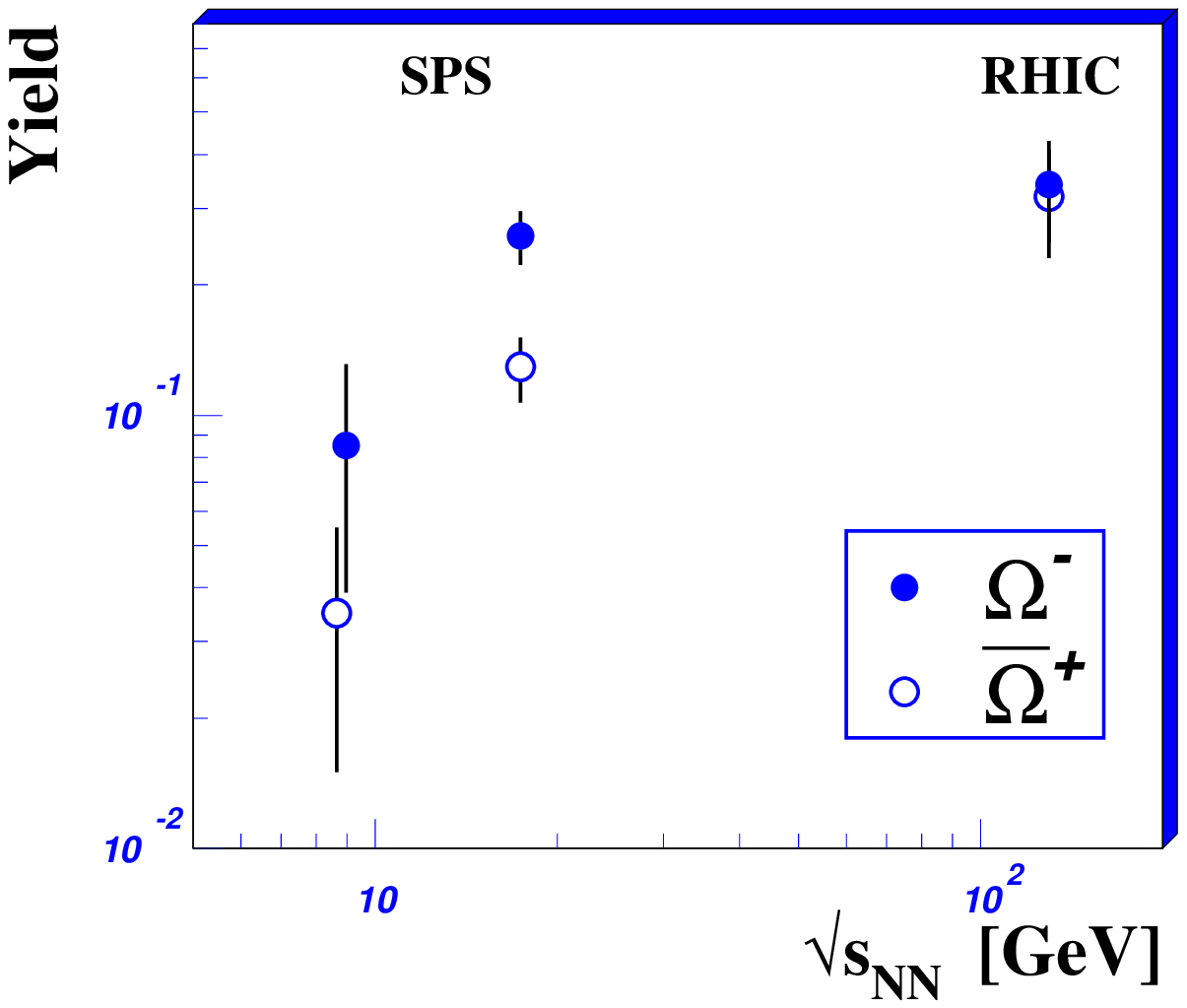}
\end{minipage}
\caption{\label{figyiecomp} K$^0_{\rm S}$ and hyperon yields
at central rapidity at SPS and RHIC energies.}
\end{figure}

NA57 has measured K$^0_{\rm S}$ and hyperon yields
as a function of centrality both in
40 and 158 $A$ GeV/$c$ Pb-Pb collisions.
Results on yields at mid-rapidity in
$\sqrt{s_{NN}}$ = 130 GeV Au-Au collisions 
have been published by the STAR Collaboration at 
RHIC~\cite{STARyieldref}. 
For comparison, we have restricted our data to the same
centrality ranges used in STAR
(most central 6\%, 5\%, 10\%, 11\%
collisions for K$^0_{\rm S}$, $\Lambda$, $\Xi$ and $\Omega$ respectively).
We show then in Figure \ref{figyiecomp} our yields per
unit rapidity at 40 $A$ GeV/$c$
($\sqrt{s_{NN}}$ = 8.8 GeV) and 158 $A$ GeV/$c$
($\sqrt{s_{NN}}$ = 17.3 GeV) togheter with those
from STAR.

\begin{figure}[h]
\centering
\includegraphics[scale=0.45]{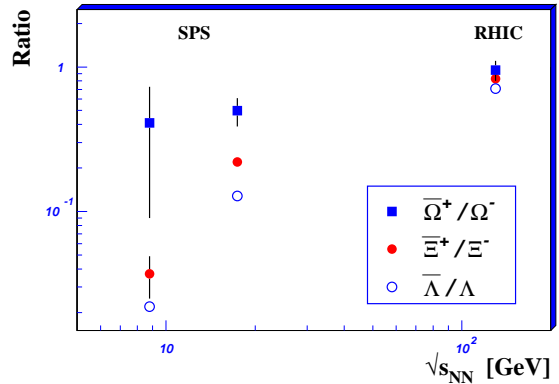}
\caption{\label{figratcomp} Comparison of antihyperon to hyperon
ratios at SPS and RHIC energies.} 
\end{figure}

The $\Lambda$ and $\Xi^-$ yields
do not vary much from SPS to RHIC,
while a clear energy dependence
is seen for all three antihyperons.
The antihyperon to hyperon ratios are plotted in
Figure \ref{figratcomp} as a function of
$\sqrt{s_{NN}}$ from SPS to RHIC~\cite{Adams2003PL}.

The large error bars on the $\overline\Omega^+$/$\Omega^-$ 
ratio are due to the restriction of the NA57 data sample
to the STAR $\Omega$ centrality range.
The ratios increase with increasing strangeness 
content of the hyperon, both at RHIC and SPS energies.
They also increase as a function
of the energy, the dependence being weaker for
particles with higher strangeness content. 
This can be understood as due to a baryon density
decrease at midrapidity with increasing energy.
\par

%As an example in Figure \ref{figyierap} 
%the $\Lambda$ and $\Xi^-$ yields are shown for
%each centrality class both for 40 and 158 $A$ GeV/$c$
%Pb-Pb collisions. 
%
%\begin{figure}[htb]
%\hspace{10mm}
%\includegraphics[scale=0.50]{../paper/y_cent_lam_t14.eps}
%\hspace{\fill}
%\begin{minipage}[t]{110mm}
%\includegraphics[scale=0.50]{../paper/y_cent_xi_t14.eps}
%\end{minipage}
%\caption{\label{figyierap} $\Lambda$ and $\Xi^-$ yields as a function
%of the event centrality for 40 and 158 $A$ GeV/$c$ Pb-Pb collisions.} 
%\end{figure}
%
%Dotted lines correspond
%to a linear increase of the yields with the number of
%wounded nucleons. 

%The $\Lambda$ and $\Xi^-$ yields grow faster than linearly
The behaviours of the yields 
with the collision centrality have been also studied.
All the yields grow faster than linearly
with the number of participants, with a steeper centrality dependence 
for the lower energy data. 
At the lower energy, the statistics does not allow
a firm conclusion for 
$\overline\Xi^+$ and $\Omega$.
The analysis of the p-Be data at 40 GeV/$c$ has allowed 
calculations of strangeness enhancements also at 
the lower energy: 
the results are reported at this conference in \cite{BrunoQM04}.

\section{Conclusions} 

Results on K$^0_{\rm S}$ and hyperon production measured
by the NA57 experiment in 
40 and 158 $A$ GeV/$c$ Pb-Pb collisions have been 
reported. The measured yields at midrapidity
have been compared
with the results of measurements carried out 
by the NA49 Collaboration: we find a 30\% systematic
discrepancy on the absolute yields, while no discrepancy 
is observed when calculating particle ratios.
\par
The energy dependence study, where yields at SPS energies
are compared with those at RHIC, shows
that $\Lambda$ and $\Xi^-$ yields per unit
rapidity stay roughly constant while a clear 
increase with energy is seen for all three
antihyperons. The antihyperon to hyperon ratios
increase with energy,
with a stronger dependence
for particles with lower strangeness content.
Such a pattern is consistent with
a decrease of the baryon density in the central region
with increasing energy.
We also observe a steeper centrality dependence
of the yields (and then of the enhancements) at lower energy.

\end{document}